\documentclass[dvips]{article}
\usepackage{icrc2011}
\usepackage{epsfig}
\usepackage{graphicx}

\title{Amplification of the Signal-to-Noise Ratio in Cosmic Ray Maps Using the Mexican Hat Wavelet Family}

\newcommand{\etal}{\MakeLowercase{\textit{et al. }}} 
\shorttitle{Rafael Alves Batista \etal Amplification of the Signal-to-Noise Ratio in Cosmic Ray Maps Using the MHWF}

\authors{Rafael Alves Batista$^{1}$, Ernesto Kemp$^{1}$, Bruno Daniel$^{1}$}

\afiliations{$^1$Instituto de F\'{i}sica ``Gleb Wataghin'' - Universidade Estadual de Campinas, Campinas-SP, Brasil, 13083-859}

\email{rab@ifi.unicamp.br}

\abstract{In this work we analyze the effect of smoothing maps containing arrival directions of cosmic rays with a gaussian kernel and kernels of the mexican hat wavelets of orders 1, 2 and 3. The analysis is performed by calculating the amplification of the signal-to-noise ratio for several background patterns (noise) and different number of events coming from a simulated source (signal) for an ideal detector capable of observing the full sky with uniform coverage. We extend this analysis for a virtual observatory with two sites, one in the northern hemisphere, the other in the southern, considering an acceptance law. }


\begin{document}
\maketitle

\section{Introduction}
 
\par There are different experiments aiming to study cosmic rays \cite{auger,hires,telescopearray,cta,magic}. They span different energy ranges and are located in different positions on the Earth, but they all share a common goal: to unveil the origin, propagation and mechanisms of acceleration of the cosmic rays, specially at the highest energies.

 \par The identification of possible astrophysical sources and the investigation of the magnetic fields which permeates the universe are studied by analyzing the arrival directions of cosmic rays. The correlation of these directions with the large structures in the universe, such as the galactic and supergalactic planes are considered large scale anisotropies. A small scale anisotropy is characterized by the association between arrival directions of cosmic rays and point sources, such as stars, distant galaxies and other kind of astrophysical objects with angular sizes small enough to be considered point-like objects. Since the sources of the cosmic rays of the highest energies are still unknown, it is an important task to disentangle genuine signals from the background.

 \section{Wavelets}
 
 \par Wavelets are localized wave-like oscillating functions belonging to the $\mathbf{L}^2$ space. When operated with a given signal, wavelets can be very useful to extract information concerning this signal, justifying its use for signal processing.

\par The continuous wavelet transform (CWT) in two dimensions may be formally written as:
\begin{equation}
 \Phi(s,\tau_1,\tau_2) = \int \int f(t,u)\Psi^{*}_{s,\tau_1,\tau_2}(t)dtdu,
\end{equation}
where $s$ ($s>0$, $s$ $\in$ $\mathbf{R}$) is the scaling factor and $\tau_1$ and $\tau_2$ ($\tau_i$ $\in$ $\mathbf{R}$) are the translation parameters. So, the CWT decomposes a function $f(t,u)$ in a basis of wavelet $\Psi_{s,\tau_1,\tau_2}(t,u)$. 

\par The function $\Psi_{s,\tau_1,\tau_2}(t,u)$ is obtained by means of scaling and translation of a so-called ``mother-wavelet'' $\Psi$:
\begin{equation}
	\Psi_{s,\tau_1,\tau_2}(t,u)=\frac{1}{\sqrt{s}} \Psi\left( \frac{t-\tau_1}{s}, \frac{u-\tau_2}{s} \right).
\end{equation}

\par The mexican hat wavelet family (MHWF) and its extension on the sphere have been widely used aiming the detection of point sources in maps of cosmic microwave background (CMB)\cite{cayon,vielva01,vielva03}, due to the amplification of the signal-to-noise ratio when going from the real space to wavelet space.

\par The MHWF is obtained by successive applications of the laplacian operator on the two-dimensional gaussian. A generic member of this family is:
\begin{equation}
	\Psi_n(\vec{x}) = \frac{(-1)^n}{2^n n!} \nabla^{2n} \phi(\vec{x}),
\end{equation}
where $\phi$ is the two-dimensional gaussian ($\phi(\vec{x}) = \frac{1}{2\pi} e^{-\vec{x}^2/2\sigma}$) and the laplacian operator is applied $n$ times.

\section{Celestial Maps}

\par Celestial maps are powerful tools to study anisotropies of cosmic rays and are ontained by pixelizing the celestial sphere taking into account the resolution of the experiment. The events map is the celestial map that represents the arrival directions of cosmic rays in the celestial sphere in a given coordinate system.

 \par Due to limitations of the detector itself, it is impossible to determine the exact arrival direction of an event. Each event detected is convolved with a probability distribution related to the angular resolution of the experiment, this is, for each event there is an associated point spreading function (PSF). Therefore, it is extremely useful to convolve the celestial maps with functions associated to the PSF of the detector, aiming to maximize the signal-to-noise ratio. Mathematically, this process of convolution (filtering\footnote{In this paper we make no distinction between the processes of convolution, smoothing and filtering. We also make no distinction between the terms kernel and filter.}) may be written as
 \begin{equation}
 	M_f (\vec{r_0}) = \alpha \int M(\vec{r}) \Phi(\vec{r},\vec{r_0})d\Omega,
 \end{equation}
  where $\alpha$ is a normalization constant, $M(\vec{r})$ is the number of cosmic ray events in the direction $\vec{r}$, $\Phi(\vec{r},\vec{r_0})$ is the kernel of the transformation and $\vec{r_0}$ is the position vector representing each point in which the integral is evaluated. In the discrete case this process is:
  \begin{equation}
  	M_f(k) = \frac{\sum_j M(j) \Phi(\vec{r_k},\vec{r_j})}{\sum_j \Phi(\vec{r_k},\vec{r_j})},
  \end{equation}
  where $M(j)$ is the number of cosmic rays associated to the pixel of index $j$ in the direction $\vec{r_j}$.

\section{Analysis Procedure}
 
 \par In the present work we have tested the capability of the MHWF to detect point sources embedded in different backgrounds. A source was simulated in the position $(l,b)$=$(320^o,30^o)$ (galactic coordinates) in the sky, with an angular size of $2^o$. The number of events in the direction of the source is 200 and the source is 10\% brighter than the background. We have also performed a similar analysis for a source with the same amplitude and 50 events coming from its direction.

\par The simulations of the background were performed according to several background patterns:
 \begin{itemize}
 	\item {\bf isotropic:} isotropic distribution of events;
 	\item {\bf dipole 1:} a dipole with excess in the galactic center $(l,b)$ = $(0^o,0^o)$, with amplitude $7\%$ with respect to the background;
 	\item  {\bf dipole 2:} a dipole with excess in the direction $(l,b)$ = $(266.5^o,-29^o)$, with amplitude $0.5\%$ with respect to the background;
 	\item {\bf sources:} several sources with different angular scales $\sigma$ and amplitudes $A$ in the directions $(l,b)$: $(0^o,0^o)$ [$\sigma=7.0^o$, $A=100\%$], $(320^o,90^o)$ [$\sigma=1.5^o$, $A=5\%$],  $(320^o,-40^o)$ [$\sigma=0.5^o$, $A=1\%$],  $(220^o,10^o)$ [$\sigma=3.0^o$, $A=5\%$],  $(100^o,-70^o)$ [$\sigma=2^o$, $A=10\%$],  $(240^o,50^o)$ [$\sigma=20^o$, $A=5\%$],  $(350^o,-80^o)$ [$\sigma=6.0^o$, $A=0.5\%$],  $(100^o,50^o)$ [$\sigma=30^o$, $A=50\%$],  $(140^o,-40^o)$ [$\sigma=4.0^o$, $A=200\%$] and  $(60^o,50^o)$ [$\sigma=3.0^o$, $A=2\%$].
 \end{itemize}
 
\par To convolve the filters with the simulated maps we have considered two different scenarios for the detector: (1) a full sky coverage with a uniform exposition, modulated only by the background pattern imposed to the simulations; (2) a detector with an acceptance law. In the last case, we considered two detectors, one in the northern hemisphere with latitude 38$^o$ N and longitude 102$^o$ W, and the other in the souther hemisphere at 36$^o$ S and 65$^o$  W. The detector on the northern hemisphere has an area seven times greater than the one in the south, so that the flux of cosmic rays is also seven times greater. Also, we have assumed that the event are detected according to a zenith angle distribution that follows $sin\theta cos\theta$, where $0^o\leq \theta \leq60^o$ is the zenith angle\cite{kalcheriess}.

 \par In order to verify the power of identification of the wavelets, we have calculated the amplification ($\lambda$) of the signal-to-noise ratio, which is given by:
 \begin{equation}
 	\lambda = \frac{w_f/\sigma_f}{w_0/\sigma_0}, 
 	\label{eq:amp}
 \end{equation}
 where $w_0$ is the value of the central pixel associated to the source  in the non-filtered source map, $w_f$ is the value of the same pixel in the filtered source map, $\sigma_0$ is the root mean square (RMS) of the non-filtered background map and $\sigma_f$ is the RMS of the filtered background map.
 
 \par
\begin{figure}[!h]
	\includegraphics[scale=0.43]{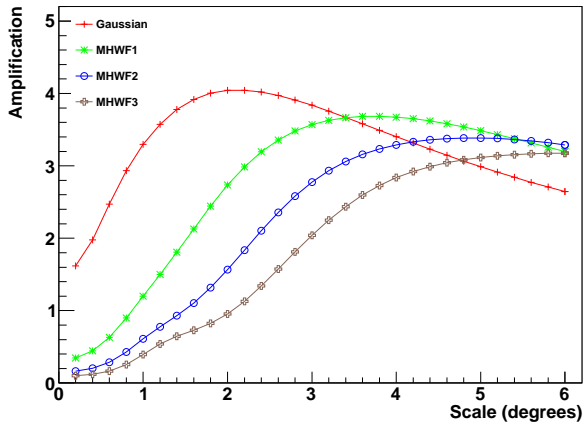}
	\caption{Amplifications for a source embedded in an isotropic background.}
	\label{fig:iso-fs}
\end{figure}

 According to Gonz{\'a}lez-Nuevo \etal\cite{mexicanhat}, the maximum amplification of the signal-to-noise ratio for a source with angular size $\gamma_0$ embedded in a white noise\footnote{White noise has the property of being homogeneous and isotropic, with a uniform power spectrum.} background is obtained by convolving the maps with a gaussian kernel with dispersion $\gamma=\gamma_0$. This shall be used as a reference to check the consistency of our results and can be seen in figure \ref{fig:iso-fs}.

\par
\begin{figure}[!h]
	\includegraphics[scale=0.43]{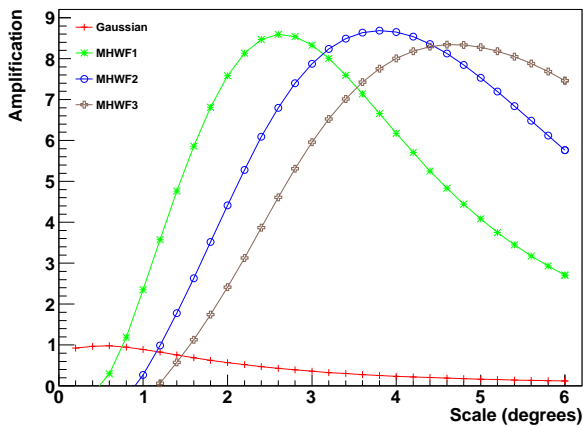}
	\caption{Amplifications for a source embedded in an isotropic background considering the acceptance of the detector.}
	\label{fig:iso}
\end{figure}

 The same reasoning can be applied for a source embedded in an isotropic background with the acceptance of the detector. However, as shown in figure \ref{fig:iso}, the gaussian kernel is not the filter that provides the best amplification.

\par For each case of signal and background we have plotted graphs similar to the ones shown in figures \ref{fig:iso-fs} and \ref{fig:iso}. 

\par
\begin{figure*}[!ht]
	\includegraphics[scale=0.83]{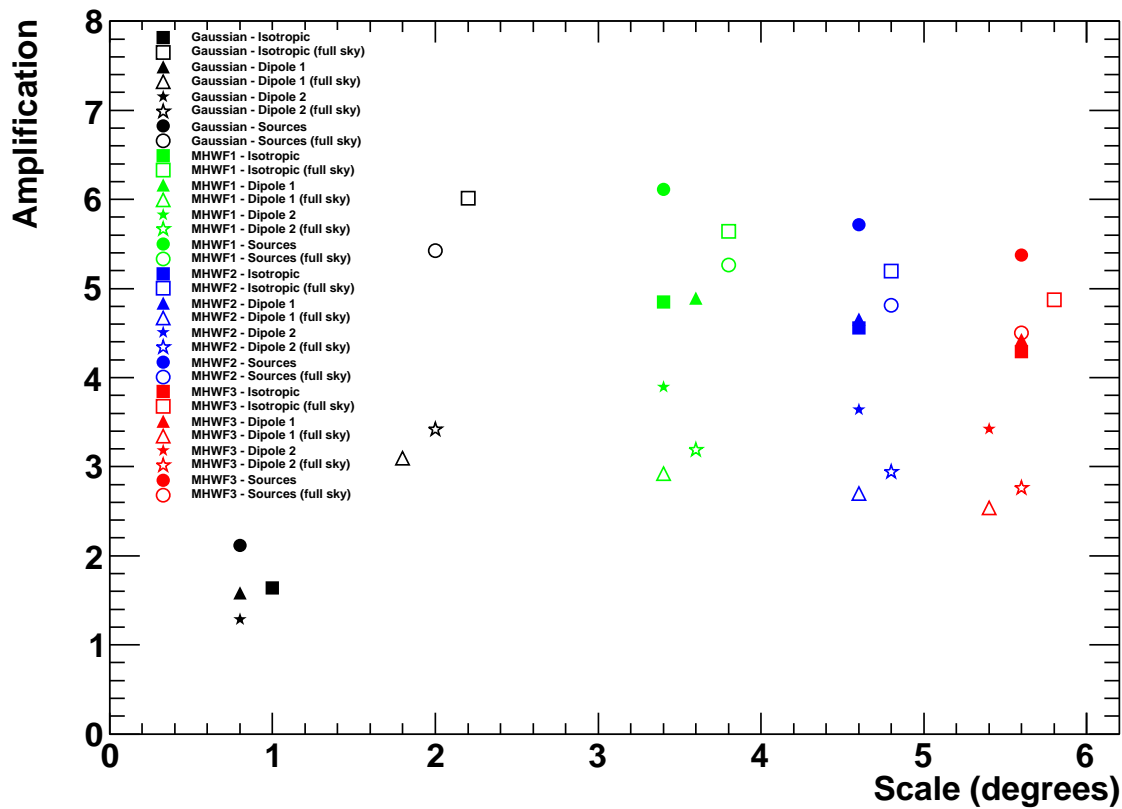}
	\caption{Maximum amplifications for each filter for all the simulated background patterns.}
	\label{fig:amplifications}
\end{figure*}

 In figure \ref{fig:amplifications} it shown the maximum value of the amplification for each filter for the different background patterns, in the case of 800,000 events from the background and 200 from the source. For all the other cases, the analysis is similar and we can draw the same conclusions. It is interesting to notice that the maximum amplification for the MHWF kernels are achieved approximately at the same scale, independently of the background pattern and the existence of an acceptance, whereas the gaussian filter is strongly dependent on the coverage.

\section{Conclusions}

\par In this work we have considered a $sin\theta cos\theta$ law for the zenith angle distribution for a virtual detector with two sites, in both hemispheres, and an ideal detector observing the whole sky, with uniform acceptance. We have simulated a source with different number of events coming from it, embedded in several background patterns. We have calculated the amplifications of the signal-to-noise ratio in each case and verified that in the case of a source with angular size $\gamma_0$ embedded in a white noise background, the maximum amplification is achieved with a gaussian filter with dispersion $\gamma=\gamma_0$, as predicted. Also, in all the cases with full coverage, the gaussian had a slightly better performance than the MHWF filters. Even though we have introduced background patterns which could lead to a background with non-white spectral features, such as a dipole and several sources, the gaussian kernel provided a slightly greater amplification, but only in the case of full sky coverage of the detector\cite{eu}.

\par It is interesting to notice that the existence of an acceptance for the detector affects the power of amplification of the gaussian filter, independently of the spectral shape of the background. The amplification, in this case, is always below 2, whereas in the case of the MHWF the amplification of the signal-to-noise ratio can be greater than 6.

\par If the background presents characteristics of non-white noise, the maximum amplification of the signal-to-noise ratio can be achived by the MHWF kernels, when compared to the gaussian. Although, when using MHWF, the directional information  regarding the source is less precise. For instance, if we consider a gaussian filter and a white
noise background, the  width of the optimal gaussian filter matches exactly with the angular size of the source, thus the uncertainty on the position would arise only from the resolution of the detector rather than the analysis method used. By using MHWF, the greater the order of the wavelet, the greater will be the difference between the angular sizes of the source and the wavelet which provides the maximum amplification. Therefore, despite the gain on the power of discrimination of the sources, the accuracy on their localization decreases and this is related only to the method and not to the instrument. We conclude that is important to consider the effects of the acceptance of the detector and their impact  on the power spectrum of the background in celestial maps and consequently, on the choice of the best kernel to be used.

\par The power spectrum of the cosmic rays is not yet fully characterized, so it is important to have analysis tools that do not strongly depend on specific features of the background to amplify the signal-to-noise ratio. Since the acceptance introduces an unknown spectral shape for the background, for analysis involving the whole sky, such as blindsearch\footnote{The procedure of blindsearch is the search for astrophysical objects within a given window around an observed excess.}, the MHWF amplifies the signal-to-noise ratio more than the gaussian. However, for small scale analysis, smoothing the maps with a gaussian kernel and reducing the area of scan could provide a greater amplification, specially if the background within the window of scan can be approximated by a white noise with a uniform power spectrum.

\par A limitation of this technique is the projection of the celestial sphere to the plane. This approximation is good enough for small scales, when we use $sin\theta \approx \theta$. For $\theta > 7^o$, however, this approximation introduces bias and this method would have to be adapted to work on a spherical manifold. Wavelets on the sphere have been used on CMB studies\cite{wiaux,mcewen1,mcewen2,mcewen3} and presented good results. The next step of this work is to perform a sistematic study relating the effect of the increase in the optimal scale of each kernel with respect to the angular size of the source. Once this is achieved, a correction factor might be applied for a better constraint on the position of the source. Furthermore, we intend to extend the method to the sphere, which allow us to search not only for point sources, but also for large scale structures in maps containing arrival directions of cosmic rays.

\section{Acknowledgements}

We are grateful for the financial support of FAPESP (Funda{\c{c}}{\~a}o de Amparo {\`a} Pesquisa do Estado de S{\~a}o Paulo), CNPq (Conselho Nacional de Pesquisa e Desenvolvimento) and CAPES (Coordena{\c{c}}{\~a}o de Aperfei{\c{c}}oamento de Pessoal de N{\'i}vel Superior).

\clearpage

\end{document}